\documentclass[12pt]{iopart}
\usepackage{graphicx}
\usepackage{subfigure}
\usepackage{epstopdf}
\usepackage{color}

\usepackage[numbers,sort&compress]{natbib}
\usepackage{verbatim}

\begin{document}

\title[Effects of NTV on plasma flow evolution in the presence of RMP in a tokamak]{Effects of neoclassical toroidal viscosity on plasma flow evolution in the presence of resonant magnetic perturbation in a tokamak}

\author{Fangyuan Ma}

\address{State Key Laboratory of Advanced Electromagnetic Technology, International Joint Research Laboratory of Magnetic Confinement Fusion and Plasma Physics, School of Electrical and Electronic Engineering, Huazhong University of Science and Technology, Wuhan 430074, China}

\author{Ping Zhu}

\address{State Key Laboratory of Advanced Electromagnetic Technology, International Joint Research Laboratory of Magnetic Confinement Fusion and Plasma Physics, School of Electrical and Electronic Engineering, Huazhong University of Science and Technology, Wuhan 430074, China}
\address{Department of Nuclear Engineering and Engineering Physics, University of Wisconsin-Madison, Madison, Wisconsin 53706, USA}

\ead{zhup@hust.edu.cn}

\author{Jiaxing Liu}

\address{State Key Laboratory of Advanced Electromagnetic Technology, International Joint Research Laboratory of Magnetic Confinement Fusion and Plasma Physics, School of Electrical and Electronic Engineering, Huazhong University of Science and Technology, Wuhan 430074, China}

\vspace{10pt}


\begin{abstract}
  Effects of neoclassical toroidal viscosity (NTV) on plasma flow evolution in the presence of resonant magnetic perturbation (RMP) in a tokamak have been evaluated using a cylindrical theory model. Calculations show that the introduction of NTV has almost no effect on the flow on the resonant surface, so the locked or unlocked state on the resonant surface remains unchanged, but it impacts the rotation profile in the core region. The toroidal, poloidal, and parallel flows in the core region are slightly reduced with uniform pressure. For non-uniform pressure profiles, elevated $\beta$  enhances the global amplitude of NTV torque but suppresses that of electromagnetic (EM) torque. These two driving terms collectively maintain the locked mode state.
\end{abstract}

\clearpage
\section{Introduction}
Resonant magnetic perturbation (RMP) has been commonly adopted for plasma control in tokamaks, such as those of edge localized modes (ELM)~\cite{Suttrop_PRL11}, rotating magnetic islands~\cite{Raob_PLA13,Dingyh_NF24}, and error field corrections~\cite{Raob_RSI13}. The interaction between RMP and tokamak plasma involves the braking and acceleration of plasma flow due to the torques induced by resonant and nonresonant  plasma response and the screening effects on plasma response to the RMP field from plasma flow~\cite{Fitzpatrick_pop14,Liuyq_pop10,Huangwl_pop15}. One such torque is the neoclassical toroidal viscosity (NTV) torque, which can dissipate momentum in the low collisionality regimes~\cite{Shaing_pop03,Shaing_NF10,Zhuw_PRL06,Garofalo_PRL08}. Previous analytical model on the field-error penetration including the effects of NTV was proposed~\cite{Cole_PRL07}. The quasi-linear code MARS-Q was developed that includes electromagnetic (EM) and NTV torques from the quasi-linear plasma response in the toroidal momentum balance equation, and the NTV torque is found to play a dominant role in certain regions of plasma for RMP penetration~\cite{Liuyq_pop13}. In general, it would be more self-consistent to consider the nonlinear interaction between plasma response and flow, with the inclusion of NTV effects from such a nonlinear plasma response.

In this work, the analytic formula for NTV torque by Shaing~\cite{Shaing_pop03} is included in a one-dimensional (1D) model for nonlinear plasma response to RMPs~\cite{Zhup_pop22}, which consists of the toroidal and poloidal torque balance equations, the modified Rutherford equation and the island phase connection condition. Calculations show that the peak magnitude of NTV torque under certain parameters is comparable to or even larger than that of EM torque. On the resonant surface, the introduction of NTV does not change the locked or unlocked states. However, NTV directly affects the rotation in the core region away from the resonant surface.

The rest of this paper is organized as follows. In Sec.~\ref{sec:the}, the theoretical model of nonlinear plasma response including the NTV torque is introduced. In Secs.~\ref{sec:uni} and~\ref{sec:non-uni}, the calculation setups and the results on the NTV effects are provided. Finally, we give a summary and discussion in Sec.~\ref{sec:sum}.

\section{Theory Model for Nonlinear Plasma Response Considering NTV Effect}
\label{sec:the}
In this work, we adopt the extended nonlinear plasma response model in a cylindrical tokamak with coordinates $(r,\theta,\phi)$ or $(r,\theta,z)$~\cite{Zhup_pop22},

\begin{equation}
r\rho\frac{\partial\Delta\Omega_\phi}{\partial t}-\frac{\partial}{\partial r}\left(r\mu\frac{\partial\Delta\Omega_\phi}{\partial r}\right)=\frac{\hat{T}_z}{4\pi^2R_0^3}\delta(r-r_s)+\hat{T}_{\rm ntv},
\label{eq:tor_rot_ntv}
\end{equation}
\begin{equation}
\frac{\partial\Delta\Omega_\phi(0,t)}{\partial r}=\Delta\Omega_\phi(a,t)=0,
\label{eq:tor_bou}
\end{equation}
\begin{equation}
r^3\rho\frac{\partial\Delta\Omega_\theta}{\partial t}-\frac{\partial}{\partial r}\left(r^3\mu\frac{\partial\Delta\Omega_\theta}{\partial r}\right)=-\frac{m\hat{T_z}}{n4\pi^2R_0}\delta(r-r_s),
\label{eq:pol_rot}
\end{equation}
\begin{equation}
\frac{\partial\Delta\Omega_\theta(0,t)}{\partial r}=\Delta\Omega_\theta(a,t)=0,
\label{eq:pol_bou}
\end{equation}
\begin{equation}
\frac{\tau_R}{1.22r_s^2}\frac{dW}{dt}=\Delta_l'+\Delta_{nl}'+\Delta_c'(\frac{W_C}{W})^2\cos\varphi,
\label{eq:width}
\end{equation}
\begin{equation}
\frac{d\varphi}{dt} = \omega_s - \frac{\sqrt2 a^2}{2A\tau_R}\Delta_c'\frac{W_C^2}{W^3}\sin\varphi,
\label{eq:phase_fslip}
\end{equation}
\begin{equation}
\hat{T}_{z}=- \frac{2 \pi^{2} R_{0}}{\mu_{0}} \frac{n}{H_{s}} E_{s c} \frac{F_{s}^{\prime 2}}{\left(4 r_{s}\right)^{4}} W^2W_C^2\sin\varphi
\label{eq:T_z}
\end{equation}
where $z=R_0 \phi$, $R_0$ is the major radius of magnetic axis, $a$ is the minor radius of plasma boundary. Eq.~(\ref{eq:tor_rot_ntv}) and Eq.~(\ref{eq:pol_rot}) are the force balance equations in the toroidal and poloidal directions respectively, Eq.~(\ref{eq:tor_bou}) and Eq.~(\ref{eq:pol_bou}) are the corresponding boundary conditions. $\hat{T}_{\rm ntv}=T_{\rm{NTV}}/r$ and $T_{\rm{NTV}}$ denotes the NTV torque density and will be detailed later. $\Omega_\phi=\Omega_{\phi 0}+\Delta\Omega_\phi$ ($\Omega_\theta=\Omega_{\theta 0}+\Delta\Omega_\theta$) is the toroidal (poloidal) rotation frequency, and $\Omega_{\phi 0}$ ($\Omega_{\theta 0}$) is the initial value. Eq.~(\ref{eq:width}) is the modified Rutherford equation for the magnetic island width $W$ due to the plasma response on the resonant surface $r_s$, and $W_C$ is the equivalent island width for the external RMP. Eq.~(\ref{eq:phase_fslip}) is the free-slip connection condition for the island phase variation, where $\varphi$ is the phase difference between the resonant magnetic response at $r_s$ and the external RMP at boundary, and $\omega_{s}=-\mathbf{k}\cdot\mathbf{u}_{s}=n\Omega_{\phi s}-m\Omega_{\theta s}$, $\Omega_{\phi s}$($\Omega_{\theta s}$) is the toroidal (poloidal) rotation frequency of plasma flow $\mathbf{u}_s$ at $r_s$.

The parameters $\Delta^\prime, F_s^\prime$ and $E_{sc}$ in Eq.~(\ref{eq:width})- (\ref{eq:T_z}) are determined from the solution of the Newcomb equation,
\begin{equation}
\frac{d^2 \psi}{d r^2}+\frac{1}{H} \frac{d H}{d r} \frac{d \psi}{d r}-\frac{1}{H}\left[\frac{g}{F^2}+\frac{1}{F} \frac{d}{d r}\left(H \frac{d F}{d r}\right)\right] \psi=0
\label{eq:Newcomb_cyl}
\end{equation}
where $\psi$ is the perturbation magnetic field in the presence of RMP boundary conditions. More details on the definitions can be found in~\cite{Huangwl_pop15,Zhup_pop22,Huangwl_pop16,Fitzpatrick_NF93,Fitzpatrick_pop01}.

The NTV torque density profile can be written as
\begin{equation}
T_{\mathrm{NTV}}=-\sum_{j=i, e} C_j \sum_n \lambda_{1, n j}\left(\Omega_{\phi}-\omega_{n c, n j}\right)
\label{eq:Tntv1}
\end{equation}
where
\begin{equation}
C_j=\rho_j R_0^2 \frac{\sqrt{\epsilon} q^2 \omega_{tj}^2}{2 \sqrt{2} \pi^{3 / 2}}
\end{equation}
\begin{equation}
\omega_{n c, n j}=q\left(\Omega_{\theta}+\omega_{* j}-\omega_{* i}+\frac{\lambda_{2, n j}}{\lambda_{1, n j}} \omega_{* T j}\right)
\label{eq:omega_nc}
\end{equation}

\begin{equation}
\lambda_{l, n j} \equiv \frac{1}{2} \int_0^{\infty} I_{\kappa n, j}(x)\left(x-\frac{5}{2}\right)^{l-1} x^{\frac{5}{2}} \mathrm{e}^{-x} \mathrm{~d} x
\end{equation}
\begin{equation}
I_{\kappa n, j} \equiv \frac{\left|n\left\langle b_n\right\rangle_{\mathrm{b}}\right|_{\kappa^2=0}^2}{v_d /(2 \epsilon)} \int_0^1 4 K F\left(I_1\left|\partial_{\kappa^2} f_{1, n j}\right|\right)^2 \mathrm{~d}\kappa^2 \end{equation}
where $\rho_j$ and $\omega_{tj}$ are the mass density and transit frequency of $j$th kind of particles, respectively, and $q$ is the safety factor. The poloidal diamagnetic drifts due to pressure and temperature gradients are $\omega_{* j}$ and $\omega_{* T j}$, respectively. $I_{\kappa n, j}$ and $\lambda_{l, n j}$ represent the pitch angle and energy integrals, respectively. $b_n$ is the amplitude of the magnetic perturbation. There are approximate analytical solutions for $I_{\kappa n, j}$ in various collisionality regimes, and the connecting formula of these analytical solutions can be used to calculate $T_{\rm{NTV}}$~\cite{Sunyw_NF11,Sunyw_NF13}.

The relationship between $\psi$ and the perturbation components in the three directions under the cylindrical configuration are as follows~\cite{Fitzpatrick_pop99},
\begin{eqnarray}
& b_r^{m, n}=\frac{i \psi}{r} \\
& b_\theta^{m, n}=-\frac{m \psi^{\prime}}{m^2+n^2 \epsilon^2}+\frac{n \epsilon \mu_0 \sigma \psi}{m^2+n^2 \epsilon^2} \\
& b_z^{m, n}=\frac{n \epsilon \psi^{\prime}}{m^2+n^2 \epsilon^2}+\frac{m \mu_0 \sigma \psi}{m^2+n^2 \epsilon^2}
\label{eq:bmn_psi}
\end{eqnarray}
where, lowercase $b$ represents the magnetic perturbation,  $m (n)$ is the poloidal (toroidal) Fourier mode number of the external perturbation, the conductivity $\sigma=(m J_{0\theta}-n\epsilon J_{0 z})/(m B_{0\theta}-n\epsilon B_{0 z})$ is calculated from the equilibrium magnetic field $\mathbf{B}_{\rm eq}=(0,B_{0\theta} (r), B_{0z})$ and current $\mathbf{J}_{e\rm q}=(0,J_{0\theta} (r), J_{0z} (r))$  in the cylindrical configuration, and $\epsilon=r/R$. The perturbation magnetic field is $\delta \vec{B}=b_r^{m, n}\vec{e_r}+b_\theta^{m, n}\vec{e_\theta}+b_z^{m, n}\vec{e_z}$ and the Fourier components of the perturbed field strength $b_n$ required for NTV calculation are as follows~\cite{Sunyw_PPCF10,Sunyw_NF12,Sunyw_PPCF15}
\begin{eqnarray}
B & \equiv B_{\rm{e q}}+\delta B,\\
\delta B &=\delta_{E} B+\delta_{\xi} B =\delta \vec{B} \cdot(\vec{B} / B)+\vec{\xi} \cdot \nabla B, \\
\delta_{E} B &=B_0 \sum_{m,n} b_{m n}^E e^{\mathrm{i}(m\theta-n\phi)},\quad \delta_{\xi} B = B_0 \sum_{m,n} b_{m n}^{\xi} e^{\mathrm{i}(m\theta-n\phi)}
\label{eq:delta_E_xi}
\\
\delta B & =-B_0 \sum_{m,n} b_{mn} e^{i(m\theta-n\phi)} \\
b_{m n} & =-\left(b_{m, n}^E+b_{m, n}^{\xi}\right)\\
b_n(\theta) & =\sum_m b_{m n} e^{\mathrm{i}(m-n q) \theta}
\label{eq:bmn}
\end{eqnarray}
where superscripts $E$ and $\xi$ respectively represent the Euler term and displacement term, as given in detail in~\cite{Sunyw_PPCF10,Sunyw_NF12,Sunyw_PPCF15}. The profile of $\psi$ is calibrated using $W=\sqrt{b_r^{m,n}/(ms B_{0\theta})_{r_s}}$ with the magnetic shear $s = (rq'/q)_{r_s}$, where the prime denotes $d/dr$. This ensures that the amplitude of the magnetic field perturbation is modulated by the magnetic island width $W$ at each time step.

\section{Equilibrium with uniform plasma pressure}
\label{sec:uni}
\subsection{Equilibrium and Parameters}
\label{sec:uni_eq}
\begin{figure}[htbp]
\centering
  \subfigure{\includegraphics[width=0.79\textwidth,angle=0]{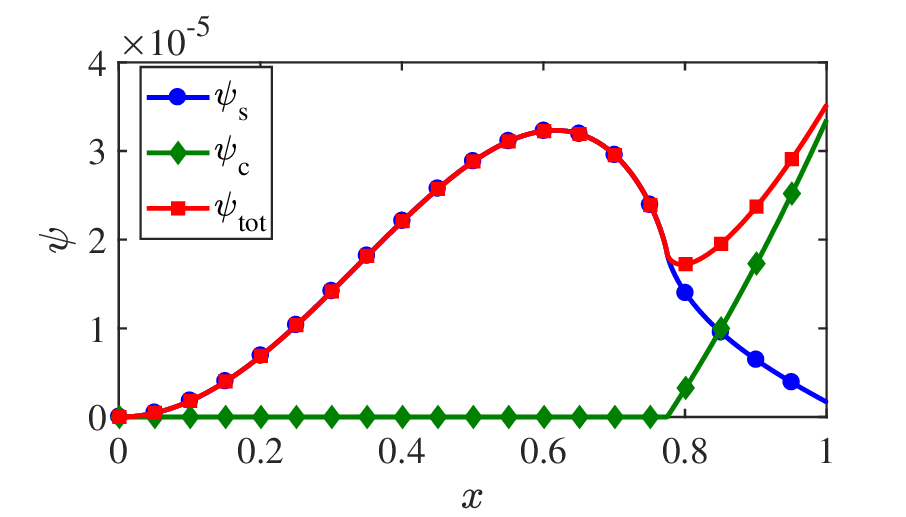}}
  \put(-80,140){(a)}

  \subfigure{\includegraphics[width=0.79\textwidth,angle=0]{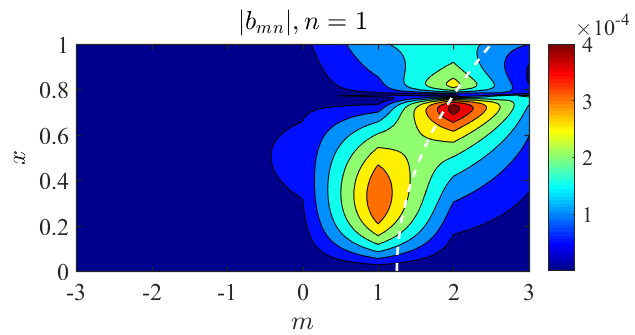}}
  \put(-280,130){\textcolor{white}{(b)}}

  \subfigure{\includegraphics[width=0.79\textwidth,angle=0]{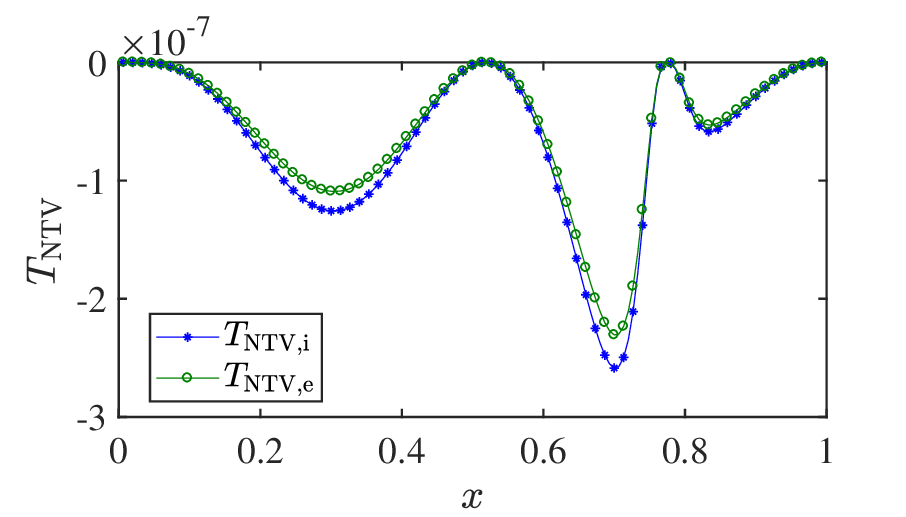}}
  \put(-280,130){(c)}
  \caption{Radial profiles of perturbed magnetic flux $\psi$, where $\psi_{tot}=\psi_s+\psi_c$, $\psi_s$ (blue) the tearing mode eigenfunction and $\psi_c$ (green) represents the external RMP field (a). Corresponding contour of the perturbed magnetic field strength $|b_{mn}|$ (b). Radial profiles of $T_{\mathrm{NTV},i}$ (blue) and $T_{\mathrm{NTV},e}$ (green) with uniform pressure (c).}
  \label{fig:psi_bnm_Tntv_ie}
\end{figure}

Consider a cylindrical tokamak with the minor radius $a=0.5{\rm{m}}$, the major radius $R_0=5{\rm{m}}$. The safety factor profile is $q(x)=1.25(1+x^2)$, where $x=r/a$ is the normalized minor radius, and $x_s=r_s/a\approx0.77460$ at the $q=2$ surface. The calculations are initialized with a radially non-uniform axisymmetric toroidal rotation $\Omega_ {\phi 0}=\Omega_0 (1-x^4)$ and $\Omega_{\theta 0}=0$ at $t=0$. The density and pressure profiles are uniform, where the number density $N = {10}^ {18}{\rm{m^{-3}}}$ or ${10} ^ {19}{\rm{m^{-3}}}$ and $\beta=\mu_0 p_0/B_0 ^ 2=0.45\%$ with $B_0=1{\rm{T}}$, $p_0$ and $B_0$ are the equilibrium values of pressure and magnetic field magnitude at the magnetic axis, respectively~\cite{Zhup_pop22}. Assuming that the ion and electron temperature are equal. In the presence of a single-helicity RMP with $m/n=2/1$, the plasma response $\psi_{tot}=\psi_s+\psi_c={\hat{\psi}}_s\Psi_s\ +{\hat{\psi}}_c\Psi_c$, where ${\hat{\psi}}_s$ and  ${\hat{\psi}}_c$ are the normalized tearing mode eigenfunction and the external RMP field, respectively. Both ${\hat{\psi}}_s$ and  ${\hat{\psi}}_c$ are evaluated using Eq.~(\ref{eq:Newcomb_cyl}) with the corresponding boundary conditions. $\Psi_s$ and $\Psi_c$ are amplitudes of the tearing mode and the RMP field which are related to the equivalent width $W$ and $W_C$, respectively. The Fourier spectrum of the perturbed field strength $b_{mn}$ of the nonlinear plasma response includes two toroidal modes (n=0,1) and the maximum poloidal number is 9 in the calculations of Eqs.~(\ref{eq:delta_E_xi})-(\ref{eq:bmn}). The $|b_{mn}|$ (Fig.~\ref{fig:psi_bnm_Tntv_ie}(b)) for the $n=1$ component is thus obtained from $\psi_{tot}$ (Fig.~\ref{fig:psi_bnm_Tntv_ie}(a)) for $W/a=0.228$ and $W_C/a=0.292$. Since both the pressure and the density are uniform,  both $\omega_ {* j}$ and $\omega_{* Tj}$ are zero, and Eq.~(\ref{eq:Tntv1}) can be reduced to
\begin{equation}
\hspace{60pt} T_{\mathrm{NTV}}=-q\omega_E \sum_{j=i, e} C_j  \sum_n \lambda_{1, n j}
\label{eq:Tntv3}
\end{equation}
where $q \omega_E=\omega_{\phi j}-q\left(\omega_{\theta j}-\omega_{* j}\right)$, and can be evaluated using $q \omega_E=\omega_{\phi i}-q\left(\omega_{\theta i}-\omega_{*i}\right)=\Omega_{\phi}-q\left(\Omega_{\theta}-\omega_{* i}\right)$ for ions. Fig.~\ref{fig:psi_bnm_Tntv_ie}(c) shows a typical  distribution of $T_{\rm{NTV,i}}$ (blue) and $T_{\rm{NTV,e}}$ (green).

\subsection{Locked state of plasma flow}
\label{sec:uni_locked}
We report a representative case of the locked state in response to RMP with $S=3\times10^5$, $Pr_m=40$ in~\cite{Zhup_pop22}, where the core toroidal rotation frequency is $\Omega_0=2\times10^2 $rad/s, the uniform number density $N=10^{18} \rm{m^{-3}}$, and $W_C/a=0.292$, which corresponds to the actual RMP amplitude value of about $4\times10^{-4}$T.

\begin{figure}
  \centering
  \subfigure{\includegraphics[width=0.9\textwidth,angle=0]{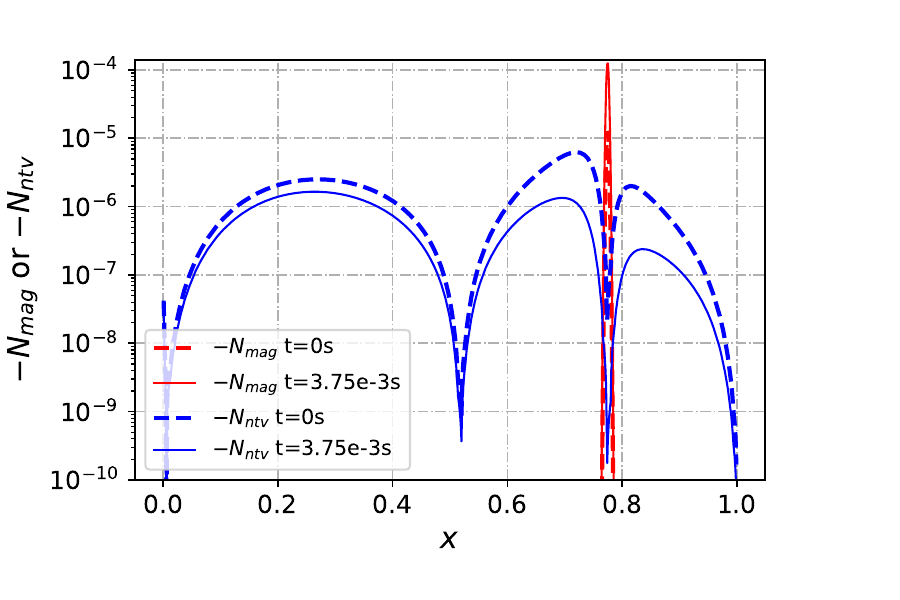}}
  \put(-100,220){(a)}

  \subfigure{\includegraphics[width=0.9\textwidth,angle=0]{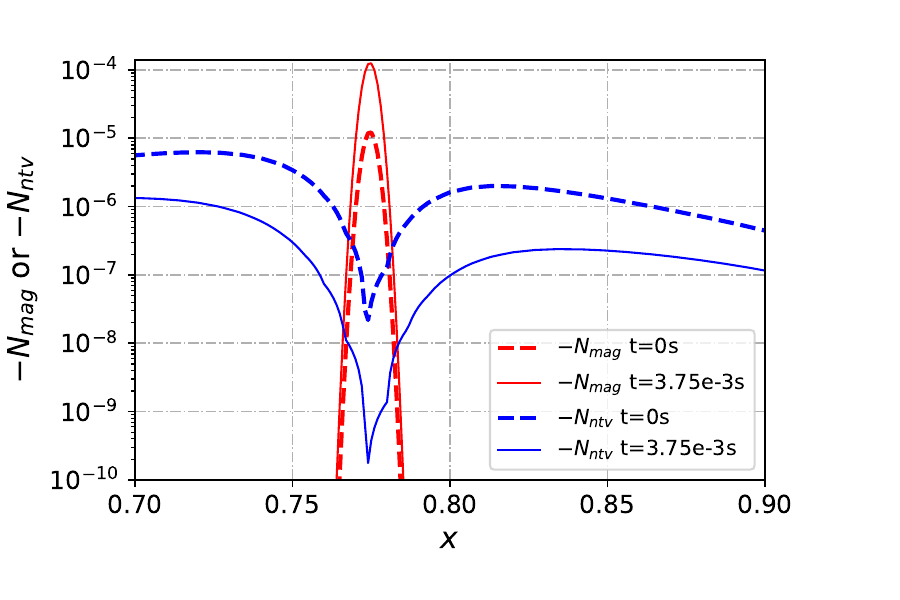}}
  \put(-100,220){(b)}
  \caption{Radial profiles of $-N_{\mathrm{mag}}$ (red) and $-N_{\mathrm{ntv}}$ (blue) at the initial (dashed) and steady state (solid) over the entire range (a) and the zoomed view near the q=2 surface (b), with the vertical axis in logarithmic scale. Here $S=3\times10^5$, $Pr_m=40$, $\Omega_0=2\times 10^2\mathrm{rad/s}$, $W_C/a=0.292$.}
  \label{fig:locked_1_N_prof}
\end{figure}

\begin{figure}[htbp]
  \centering
  \subfigure{\includegraphics[width=0.9\textwidth,angle=0]{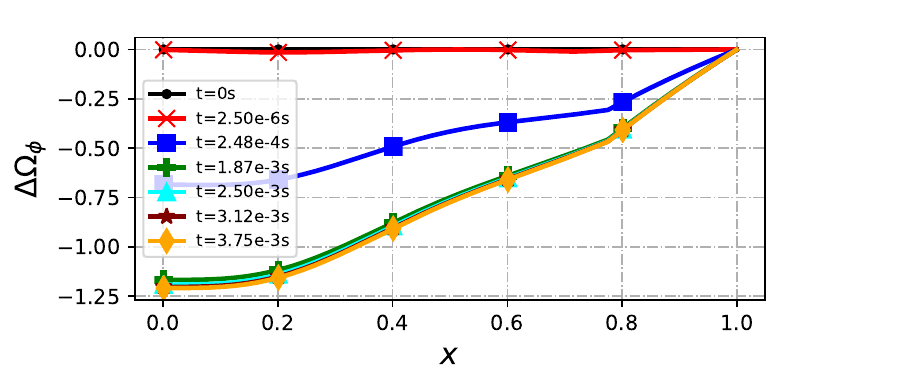}}
  \put(-105,110){(a)}

  \subfigure{\includegraphics[width=0.9\textwidth,angle=0]{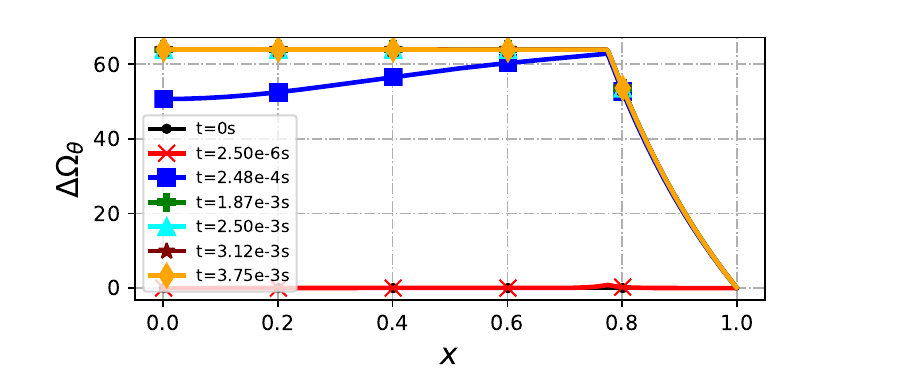}}
  \put(-105,110){(b)}

  \subfigure{\includegraphics[width=0.9\textwidth,angle=0]{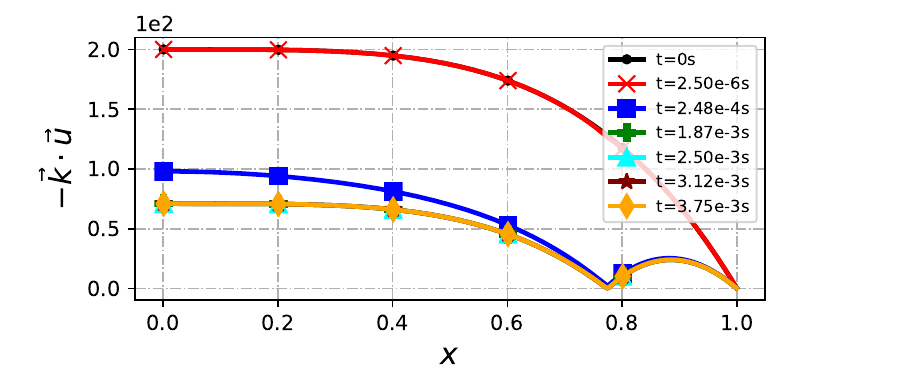}}
  \put(-320,125){(c)}

  \caption{Radial profiles of $\Delta \Omega_\phi$ (a), $\Delta \Omega_\theta$ (b) and $-\mathbf{k}\cdot\mathbf{u}$ (c) at different time slices. Here $S=3\times10^5$, $Pr_m=40$, $\Omega_0=2\times 10^2\mathrm{rad/s}$, $W_C/a=0.292$.}
  \label{fig:locked_1_ku_prof}
\end{figure}

\begin{figure}[htbp]
  \centering
  \subfigure{\includegraphics[width=0.9\textwidth,angle=0]{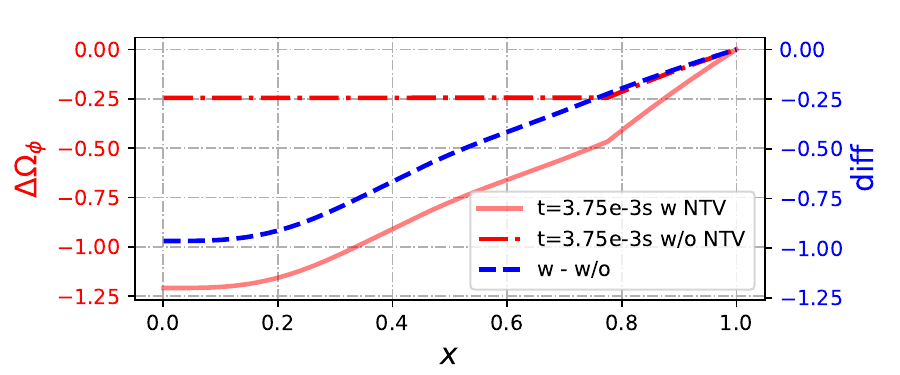}}
  \put(-105,110){(a)}

  \subfigure{\includegraphics[width=0.9\textwidth,angle=0]{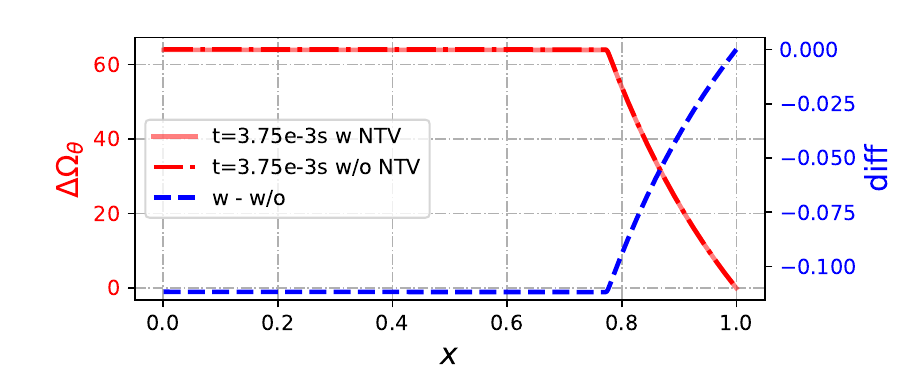}}
  \put(-110,120){(b)}

  \subfigure{\includegraphics[width=0.9\textwidth,angle=0]{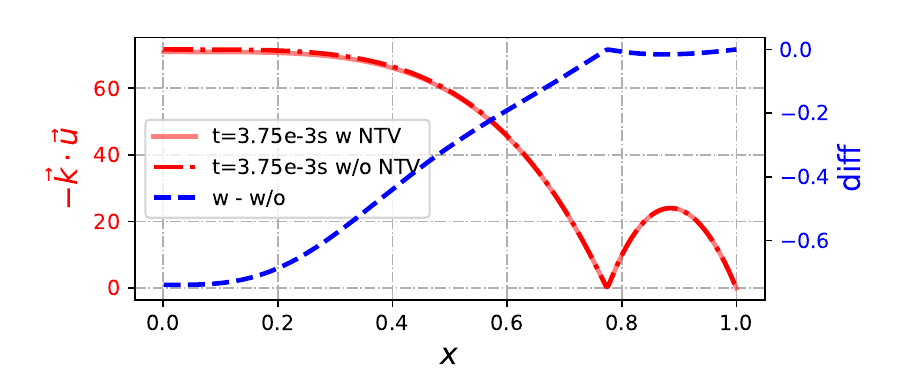}}
  \put(-105,110){(c)}

  \caption{Radial profiles of $\Delta \Omega_\phi$ (a), $\Delta \Omega_\theta$ (b) and $-\mathbf{k}\cdot\mathbf{u}$ (c) in the steady state with NTV effect (red solid) and without it (red dash-dotted) and their difference (blue dashed). Here $S=3\times10^5$, $Pr_m=40$, $\Omega_0=2\times 10^2\mathrm{rad/s}$, $W_C/a=0.292$.}
  \label{fig:locked_1_ku_prof_diff}
\end{figure}

\begin{figure}[htbp]
  \centering
  \subfigure{\includegraphics[width=0.9\textwidth,angle=0]{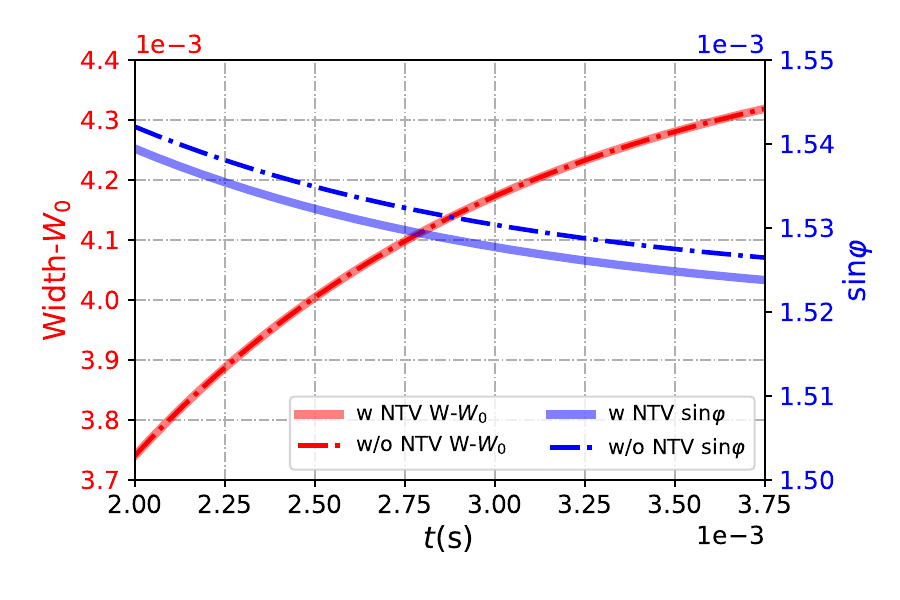}}

  \caption{ The width of magnetic island (red) and $\sin\varphi$ (blue) as function of time with NTV effect (solid), without it (dash-dotted). Here $S=3\times10^5$, $Pr_m=40$, $\Omega_0=2\times 10^2\mathrm{rad/s}$, $W_C/a=0.292$ and $W_0$=0.11.}
  \label{fig:locked_1_evo}
\end{figure}

Fig.~\ref{fig:locked_1_N_prof} shows the comparison between the EM torque  $N_{\mathrm{mag}}=\hat{T}_z\delta(r-r_s)/({4\pi^2R_0^3})$ (red) and NTV torque $N_{\mathrm{ntv}}=\hat{T}_{\rm ntv}$ (blue) at the initial (dashed line) and the steady states (solid line) . The EM torque only exists near the $q=2$ resonant surface, whereas the NTV torque is distributed throughout most of the radial region and remains almost zero on the resonant surface. The NTV torque is smaller than the initial value in the entire region near the steady state, since $\omega_E$ is the maximum initially, and gradually decreases with time. Changes in other parameters can also have different effects on the profiles of NTV torque, but in this case, the change in $\omega_E$ has the most significant impact. Figs.~\ref{fig:locked_1_ku_prof}(a)-(c) represent radial profiles of $\Delta\Omega_\phi$, $\Delta\Omega_\theta$, and $-\mathbf {k} \cdot \mathbf{u}$ respectively at different time slices with NTV under uniform pressure with the inclusion of the NTV effect. Their steady-state solutions (red solid lines) are compared with the steady-state solutions without the NTV effect (red dash-dotted lines), as shown in Figs.~\ref{fig:locked_1_ku_prof_diff}(a)-(c), respectively, where the blue dashed lines represent the absolute errors between the two. It can be seen that the introduction of NTV effect has a direct and significant impact on the toroidal rotation. The EM torque and NTV torque in Eq.~(\ref{eq:tor_rot_ntv}) have the same direction and are negative, resulting in a decrease in $\Delta \Omega_\phi$, especially in the core region. Note that the presence of viscous terms containing $\mu$ in Eq.~(\ref{eq:tor_rot_ntv}) will cause the influence of those torques to spread towards the core. Although the NTV term is not included in the poloidal rotation equation Eq.~(\ref{eq:pol_rot}), it will indirectly modify the poloidal rotation via $\hat{T}_{z}$. The indirect effect of the NTV on poloidal rotation is a slight decrease in $\Omega_\theta$, which indicates that the absolute value of the EM torque term on the right side of the Eq.~(\ref{eq:pol_rot}) decreases after the introduction of NTV. For the convenience of clarification, Eq.~(\ref{eq:T_z}) can be simplified as $\hat{T}_z \sim W^2 \sin\varphi$, while the remaining physical quantities remain constant during the evolution of the equation. Fig.~\ref{fig:locked_1_evo} shows the evolution of island width (red) and $\sin\varphi$ (blue) over time with (solid line) and without (dash-dotted line) NTV effect. The decrease in $\sin \varphi$ is much more significant than the slight increase in $W$, ultimately resulting in a decrease in the absolute value of ${\hat {T}}_z$.

In the $\phi$ direction, the EM torque is negative, while in the $\theta$ direction, it is positive. Under the combined influence of toroidal and poloidal rotation, the core of $-\mathbf {k} \cdot \mathbf {u}$ is lower, but the $-\mathbf {k} \cdot \mathbf{u_s}$ on the resonant surface remains almost unchanged (Fig.~\ref{fig:locked_1_ku_prof_diff}(c)), that is to say, the introduction of NTV effect does not change the locked state on the resonant surface.

\subsection{Unlocked state of plasma flow}
\label{sec:uni_unlocked}
\begin{figure}[htbp]
  \centering
  \subfigure{\includegraphics[width=0.9\textwidth,angle=0]{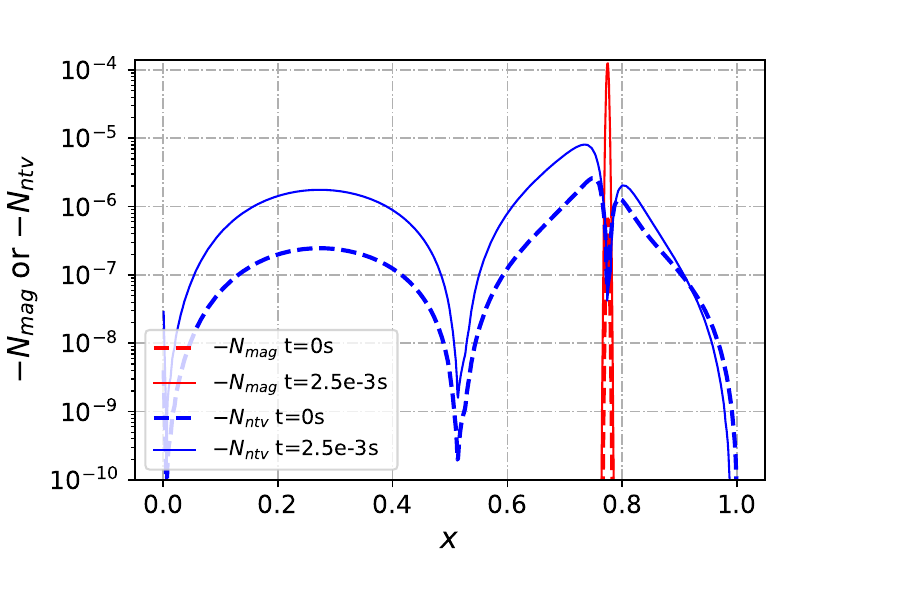}}
  \put(-100,220){(a)}

  \subfigure{\includegraphics[width=0.9\textwidth,angle=0]{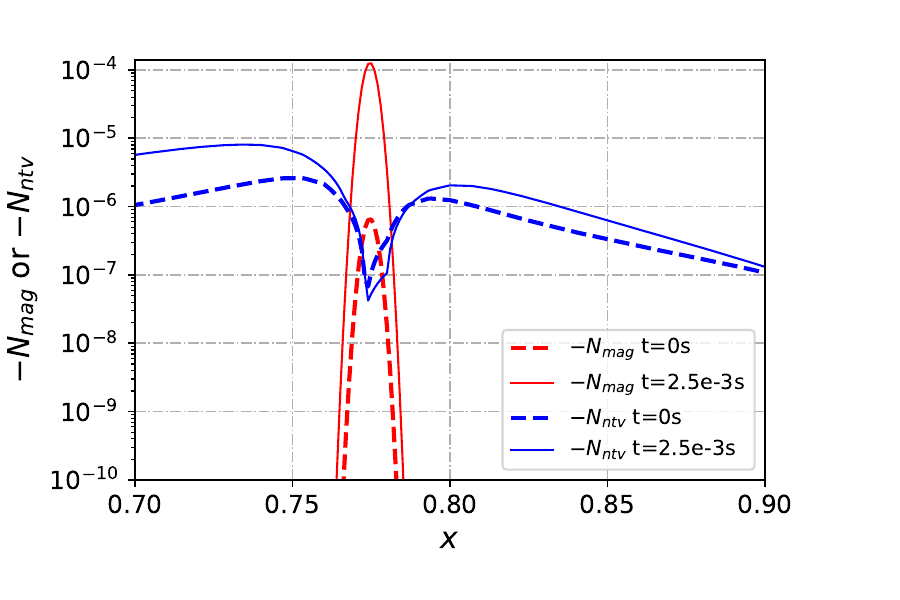}}
  \put(-100,220){(b)}

  \caption{Radial profiles of $-N_{\mathrm{mag}}$ (red) and $-N_{\mathrm{ntv}}$ (blue) at the initial (dashed) and steady state (solid) over the entire range (a) and the zoomed view near the q=2 surface (b), with the vertical axis in logarithmic scale. Here $S=2.44\times10^3$, $Pr_m=1$, $\Omega_0=2\times 10^2 \mathrm{rad/s}$, $W_C/a=0.146$.}
  \label{fig:unlocked_1_N_prof}
\end{figure}

\begin{figure}[htbp]
  \centering
  \subfigure{\includegraphics[width=0.9\textwidth,angle=0]{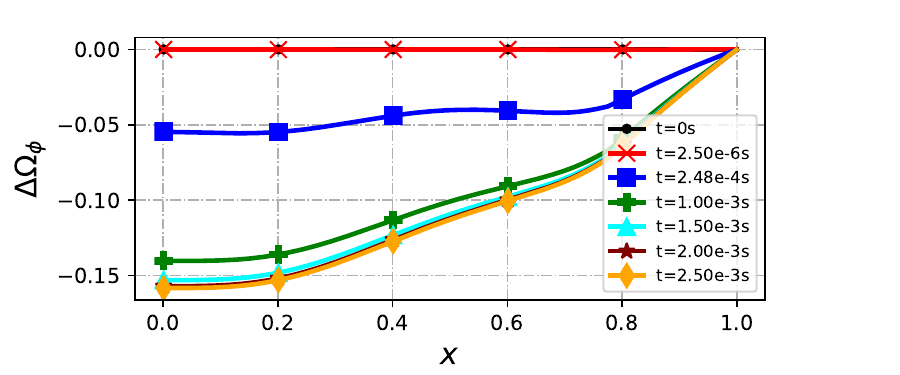}}
  \put(-320,125){(a)}

  \subfigure{\includegraphics[width=0.9\textwidth,angle=0]{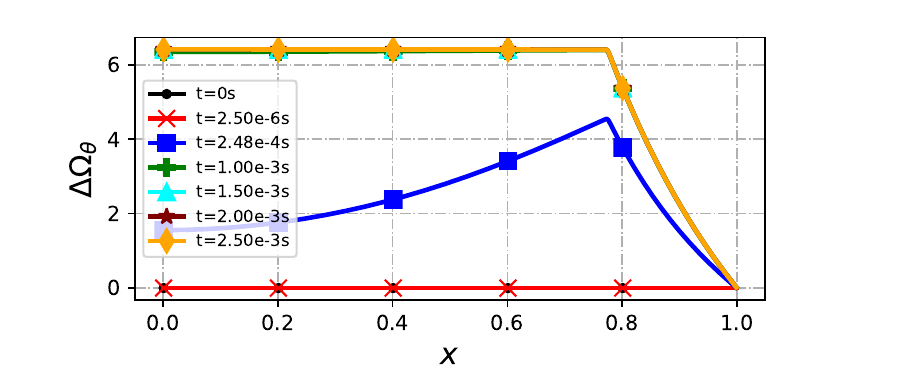}}
  \put(-105,125){(b)}

  \subfigure{\includegraphics[width=0.9\textwidth,angle=0]{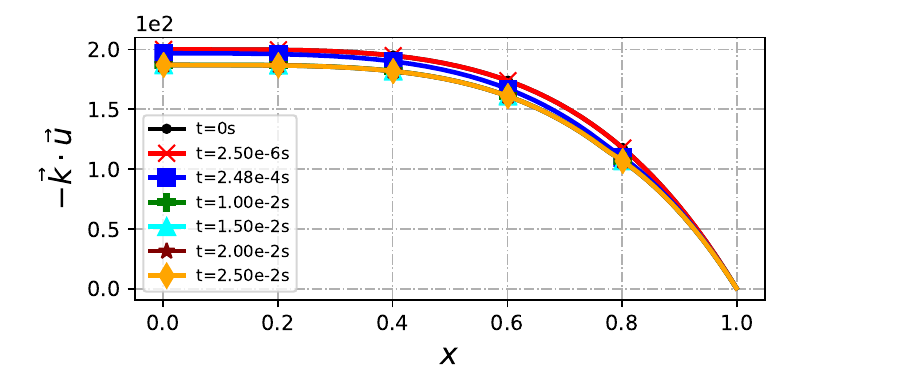}}
  \put(-105,125){(c)}

  \caption{Radial profiles of $\Delta \Omega_\phi$ (a), $\Delta \Omega_\theta$ (b) and $-\mathbf{k}\cdot\mathbf{u}$ (c) at different time slices. Here  $S=2.44\times10^3$, $Pr_m=1$, $\Omega_0=2\times 10^2 \mathrm{rad/s}$, $W_C/a=0.146$.}
  \label{fig:unlocked_1_ku_prof}
\end{figure}

\begin{figure}[htbp]
  \centering
  \subfigure{\includegraphics[width=0.9\textwidth,angle=0]{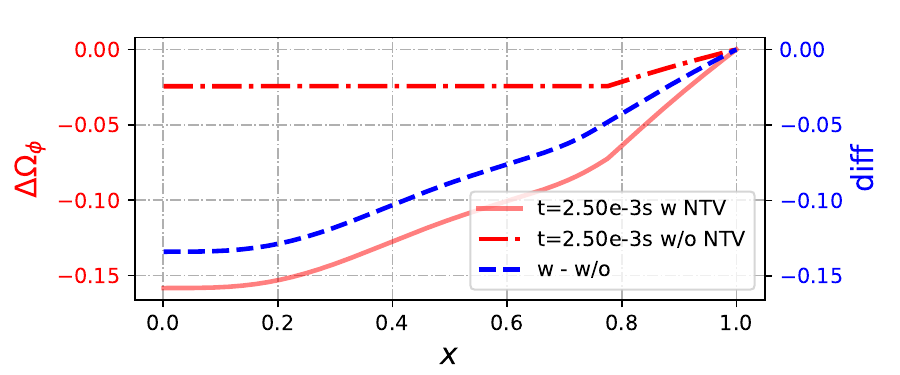}}
  \put(-90,120){(a)}

  \subfigure{\includegraphics[width=0.9\textwidth,angle=0]{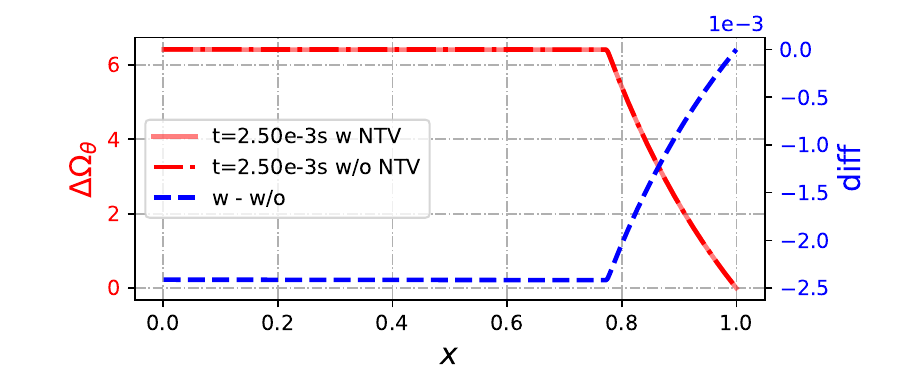}}
  \put(-110,120){(b)}

  \subfigure{\includegraphics[width=0.9\textwidth,angle=0]{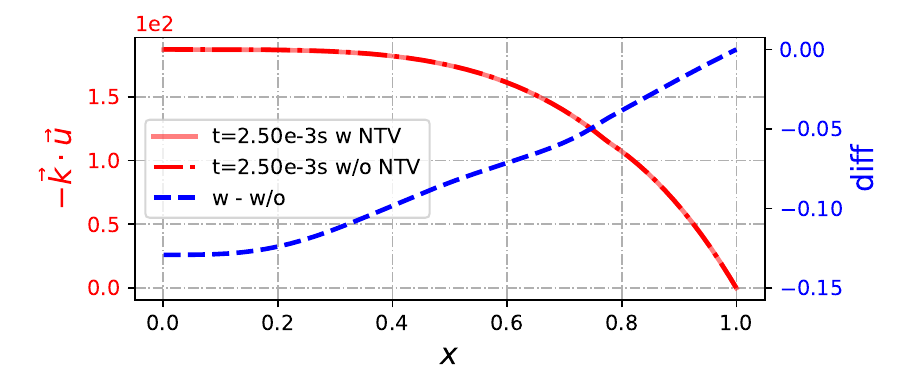}}
  \put(-90,110){(c)}

  \caption{Radial profiles of $\Delta \Omega_\phi$ (a), $\Delta \Omega_\theta$ (b) and $-\mathbf{k}\cdot\mathbf{u}$ (c) in the steady state with NTV effect (red solid) and without it (red dash-dotted) and their difference (blue dashed). Here  $S=2.44\times10^3$, $Pr_m=1$, $\Omega_0=2\times 10^2 \mathrm{rad/s}$, $W_C/a=0.146$.}
  \label{fig:unlocked_1_ku_prof_diff}
\end{figure}

For a sufficiently weak RMP amplitude, the nonlinear plasma response to RMP can reach a steady state where the plasma flow remains finite on the resonant flux surface, which is referred to as the unlocked state of plasma flow.
Here, we also report a representative case of the unlocked state with $W_C/a=0.146$, $N=10^{19}\ \rm{m^{-3}}$, $S=2.44\times10^3$ and $Pr_m=1$. The initial toroidal rotation frequency is set identical to that of the locked state case, i.e., $\Omega_0=200\ \rm{rad/s}$.
The NTV torque in the steady state is larger than that in the initial state (Fig.~\ref{fig:unlocked_1_N_prof}). This is because $\omega_E$ remains nearly unchanged in this case, whereas the magnetic perturbation $b_{nm}$ grows as the magnetic island expands.
The introduction of NTV effect slightly decreases the entire profile of $-\mathbf {k} \cdot \mathbf {u}$ (Fig.~\ref{fig:unlocked_1_ku_prof}(c)), and the change on the resonant surface is comparable to that in other regions (blue dashed line in Fig.~\ref{fig:unlocked_1_ku_prof_diff}(c)). The unlocked state is not altered by NTV.

\section{Equilibrium with non-uniform plasma pressure}
\label{sec:non-uni}
\subsection{Equilibrium and Parameters}
\label{sec:non-uni_eq}
In this section, a non-uniform pressure profile $p(x)=p_0(1-x^2)$ is adopted. Note that $p_0$ is actually the maximum pressure here. Other parameter setups and assumptions are the same as those in Sec.~\ref{sec:uni_eq}.

Calculations have shown that if there is no pressure gradient, $\omega_{*, j}$ and $\omega_{*T, j}$ will always be zero, and the magnitude of $N_{\mathrm{ntv}}$ is relatively small. Now the gradients are introduced, $\omega_{*, j}$ and $\omega_{*T, j}$ are not zero. When the number density $N={10}^{18}\sim{10}^{19}{\rm{m^{-3}}}$, $\beta=0.5\%\sim1\%$, $N_{\mathrm{ntv}}$ can be comparable to the magnitude of the $N_{\mathrm{mag}}$. As $\beta$ further increases, the peak value of $N_{\mathrm{ntv}}$ will be much larger than that of $N_{\mathrm{mag}}$.

\subsection{Locked state of plasma flow}
\label{sec:non-uni_locked}
\begin{figure}[htbp]
\centering
  \subfigure{\includegraphics[width=0.9\textwidth,angle=0]{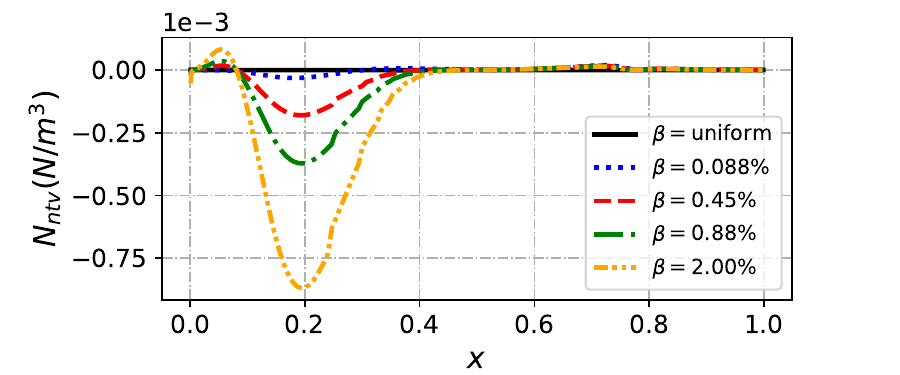}}
  \put(-100,120){(a)}

  \subfigure{\includegraphics[width=0.9\textwidth,angle=0]{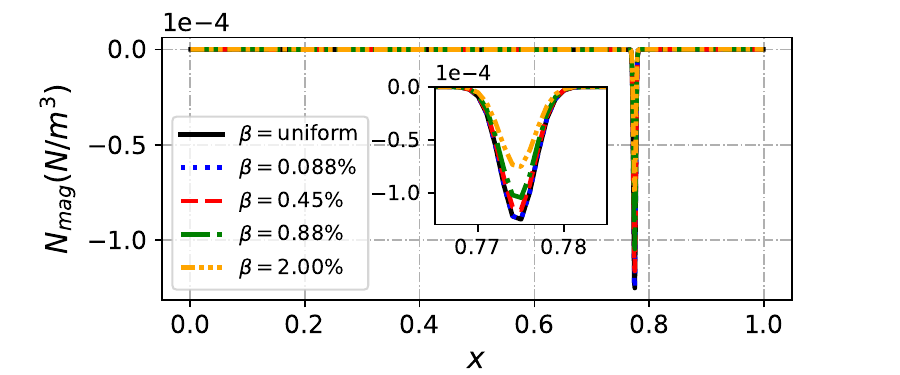}}
  \put(-100,120){(b)}

  \subfigure{\includegraphics[width=0.9\textwidth,angle=0]{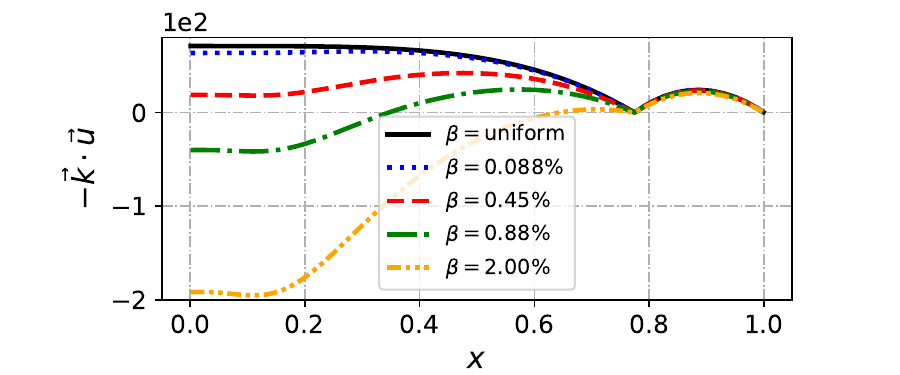}}
  \put(-100,135){(c)}

  \caption{Radial profiles of $N_{\mathrm{ntv}}$ (a), $N_{\mathrm{mag}}$ (b) and $-\mathbf{k}\cdot\mathbf{u}$ (c) for various $\beta$ in the steady state under non-uniform pressure. Here $S=3\times10^5$, $Pr_m=40$, $\Omega_0=2\times 10^2 \mathrm{rad/s}$, $W_C/a=0.292$.}
  \label{fig:locked_1_dpdr_only_NTV_prof}
\end{figure}

\begin{figure}[htbp]
\centering
  \subfigure{\includegraphics[width=0.9\textwidth,angle=0]{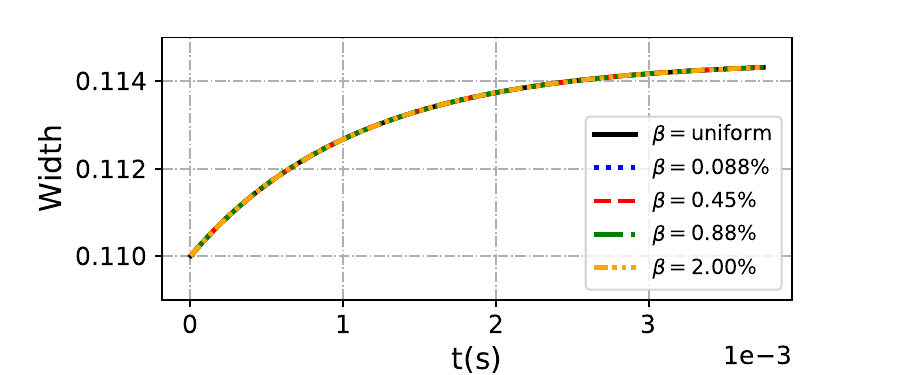}}
  \put(-300,135){(a)}

  \subfigure{\includegraphics[width=0.9\textwidth,angle=0]{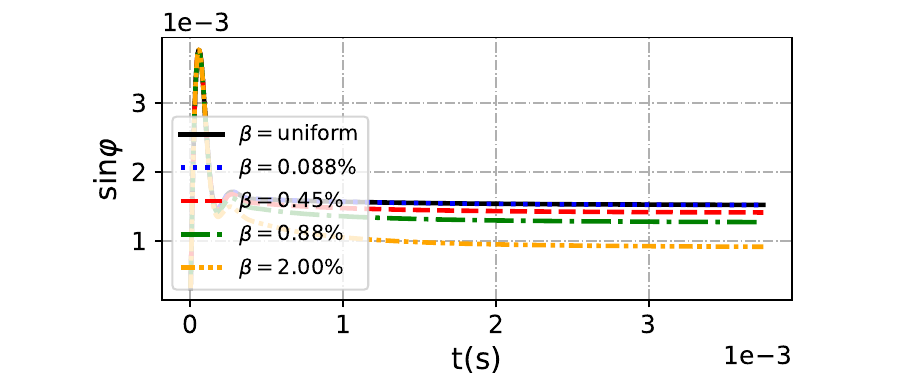}}
  \put(-300,135){(b)}

  \subfigure{\includegraphics[width=0.9\textwidth,angle=0]{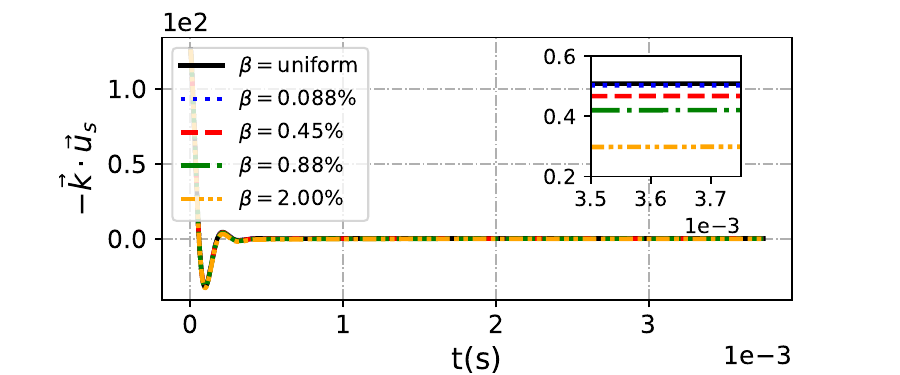}}
  \put(-230,135){(c)}

  \caption{The width of magnetic island (a), $\sin\varphi$ (b) and  $-\mathbf{k}\cdot\mathbf{u}_s$ (c) as function of time for various $\beta$ in the steady state under non-uniform pressure. Here $S=3\times10^5$, $Pr_m=40$, $\Omega_0=2\times 10^2 \mathrm{rad/s}$, $W_C/a=0.292$.}
  \label{fig:locked_1_dpdr_only_NTV_evo}
\end{figure}

Figs.~\ref{fig:locked_1_dpdr_only_NTV_prof}(a)-(c) present the radial profiles of $N_{\mathrm{ntv}}$, $N_{\mathrm{mag}}$, and $-\mathbf{k}\cdot\mathbf{u}$ at steady state under various $\beta$ for the case with $S=3 \times 10^5$, $Pr_m=40$. And the corresponding magnetic island width $W$, $\sin\varphi$, and $-\mathbf{k}\cdot\mathbf{u}_s$ for this case as functions of time are shown in Figs.~\ref{fig:locked_1_dpdr_only_NTV_evo}(a)-(c). Note that the black solid lines correspond to results obtained under uniform pressure conditions in Sec.~\ref{sec:uni}.
Notably, the magnitude of $N_{\mathrm{ntv}}$ (Fig.~\ref{fig:locked_1_dpdr_only_NTV_prof}(a)) increases globally with the increase of $\beta$. The magnetic island width $W$ remains largely insensitive to $\beta$ (Fig.~\ref{fig:locked_1_dpdr_only_NTV_evo}(a)), and the variation in $N_{\mathrm{ntv}}$ magnitude primarily arises from changes in the pressure gradient. The $-\mathbf{k}\cdot\mathbf{u}$ profile in the core region (Fig.~\ref{fig:locked_1_dpdr_only_NTV_prof}(c)) is directly related to the distribution of $N_{\mathrm{ntv}}$ shown in Fig.~\ref{fig:locked_1_dpdr_only_NTV_prof}(a). The $-\mathbf{k}\cdot\mathbf{u}_s$ on the resonant surface approaches zero in the steady state (Fig.~\ref{fig:locked_1_dpdr_only_NTV_prof}(c) and Fig.~\ref{fig:locked_1_dpdr_only_NTV_evo}(c)), which means that the locked state is sustained across various $\beta$.

The magnitude of $N_{\mathrm{mag}}$ decreases with $\beta$ (Fig.~\ref{fig:locked_1_dpdr_only_NTV_prof}(b)) and the magnitude of this decrease is consistent with the variation of $\sin\varphi$ (Fig.~\ref{fig:locked_1_dpdr_only_NTV_evo}(b)) (note that $\hat{T}_z \propto W^2 \sin\varphi$ while $W$ is insensitive to $\beta$). Given the approximation $\lim_{\varphi \to 0} \sin \varphi \approx \varphi$, the temporal evolution of $\varphi$ is thus consistent with the trend shown in Fig.~\ref{fig:locked_1_dpdr_only_NTV_evo}(b). As $\varphi$ approaches zero, $\sin\varphi$ varies sharply while $\cos\varphi$ remains nearly constant. This explains why $\hat{T}_z$ and $d\varphi/dt$ vary significantly with $\beta$, while $W$ remains largely unchanged. The influence of $\beta$ on $\varphi$ originates from the $\omega_s$ in Eq.~(\ref{eq:phase_fslip}). As illustrated in the zoomed view of Fig.~\ref{fig:locked_1_dpdr_only_NTV_evo}(c), when $\beta$ transitions from a uniform to a 2\% non-uniform distribution, $-\mathbf{k}\cdot\mathbf{u}_s$ in the steady state decreases by 40\%.

By contrast, as $\beta$ increases, $N_{\mathrm{ntv}}$ exhibits a global increase in magnitude. The introduction of NTV slightly reduces the parallel flow $\omega_s$ at the resonant surface, and this effect becomes increasingly pronounced with increasing $\beta$. Moreover, this leads to a reduction in the phase difference $\varphi$ via the island phase equation Eq.~(\ref{eq:phase_fslip}), thereby diminishing the magnitude of $N_{\mathrm{mag}}$. The competitive interplay between $N_{\mathrm{ntv}}$ and $N_{\mathrm{mag}}$ ultimately sustains the locked state.

\subsection{Unlocked state of plasma flow}
\label{sec:non-uni_unlocked}
\begin{figure}[htbp]
\centering
  \subfigure{\includegraphics[width=0.9\textwidth,angle=0]{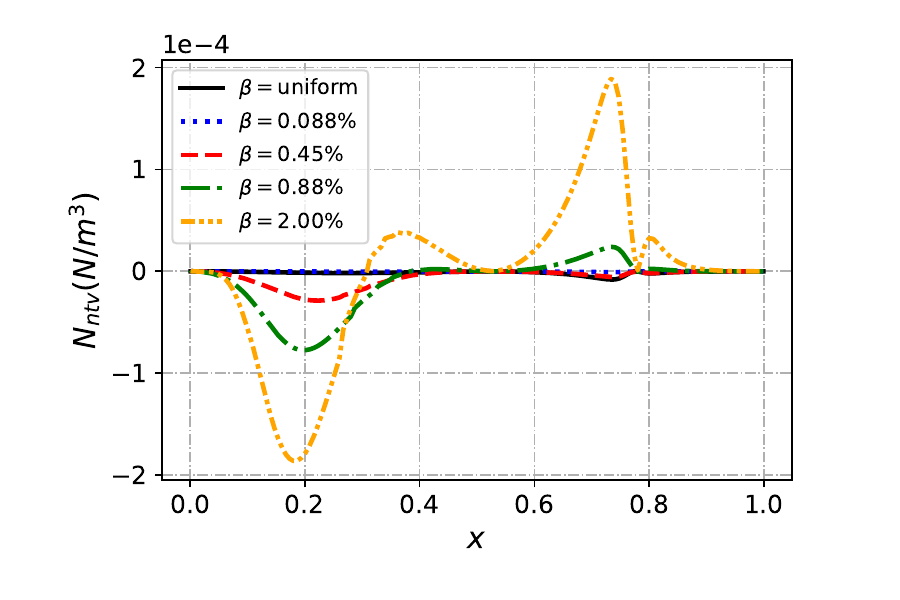}}
  \put(-100,200){(a)}

  \subfigure{\includegraphics[width=0.9\textwidth,angle=0]{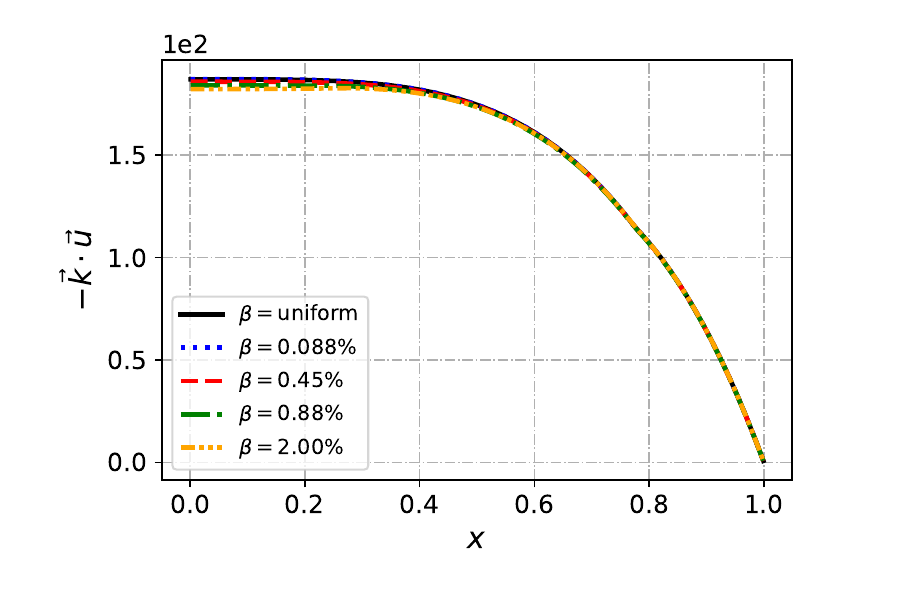}}
  \put(-100,220){(b)}

  \caption{Radial profiles of $N_{\mathrm{ntv}}$ (a) and $-\mathbf{k}\cdot\mathbf{u}$ (b) for various $\beta$ in the steady state under non-uniform pressure. Here $S=2.44\times10^3$, $Pr_m=1$, $\Omega_0=2\times 10^2 \mathrm{rad/s}$, $W_C/a=0.146$.}
  \label{fig:unlocked_1_dpdr_only_NTV_prof}
\end{figure}

Similarly, we perform calculations for the unlocked case mentioned in Sec.~\ref{sec:uni_unlocked} under non-uniform pressure conditions. Calculations show that the magnitude of $N_{\mathrm{ntv}}$ in the core region increases with $\beta$. Around the $q=2$ resonant surface, $N_{\mathrm{ntv}}$ exhibits a negative direction under uniform and lower $\beta$ , but shifts to positive at higher $\beta$, with its magnitude varying non-monotonically (Fig.~\ref{fig:unlocked_1_dpdr_only_NTV_prof}(a)). The evolution of $W$ and $\sin\varphi$ is almost unaffected by $\beta$, and the amplitude of $N_{\rm mag}$ is naturally nearly independent of $\beta$. The value of $-\mathbf{k}\cdot\mathbf{u}$ in the core region decreases with increasing $\beta$, while $-\mathbf{k}\cdot\mathbf{u}_s$ at the $q=2$ resonant surface remains almost unchanged under various $\beta$, thus consistently maintaining the unlocked state (Fig.~\ref{fig:unlocked_1_dpdr_only_NTV_prof}(b)).

\subsection{Effects of temperature profile flattening}
\label{sec:T-flatten}
\begin{figure}[htbp]
\centering
  \subfigure{\includegraphics[width=0.9\textwidth,angle=0]{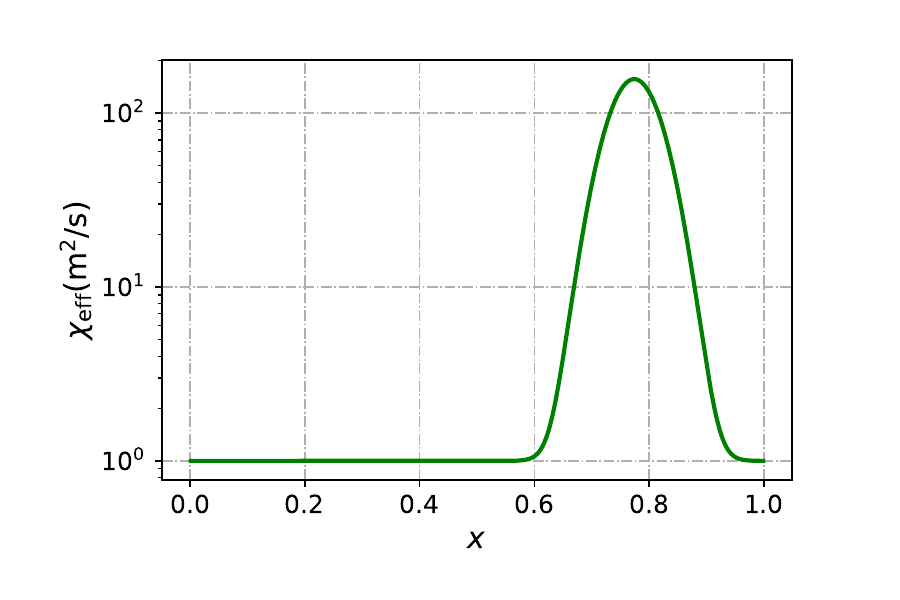}}
  \put(-80,220){(a)}

  \subfigure{\includegraphics[width=0.9\textwidth,angle=0]{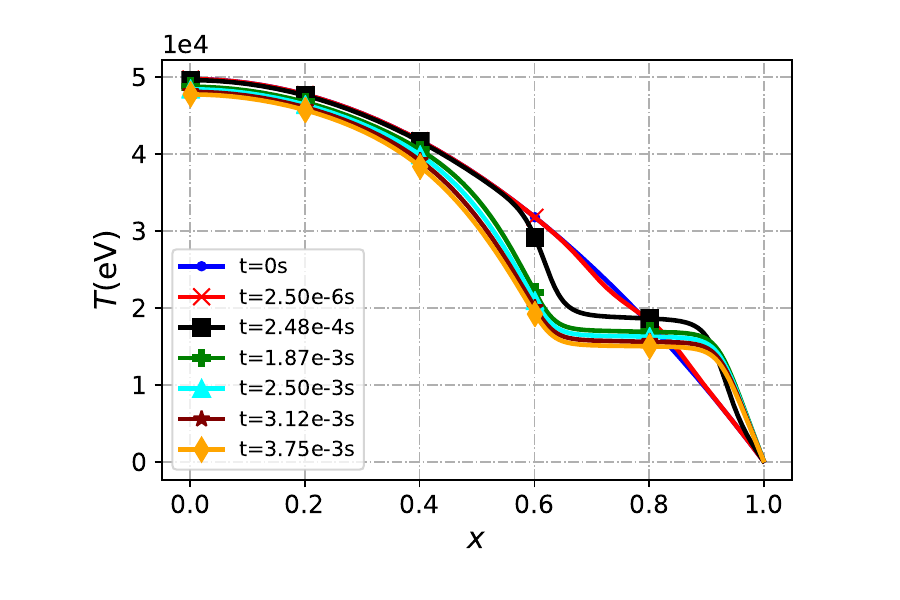}}
  \put(-80,220){(b)}

  \caption{Radial profile of $\chi_{\rm eff}$ at $t=0$ with logarithmic scale for vertical axis (a) and temperature $T$ profiles with $\beta=2\%$ at different time slices (b). }
  \label{fig:T_flat}
\end{figure}

In the previous section, the pressure or temperature profile does not evolve, whether in the case of uniform or non-uniform pressure. However, anisotropic heat transport occurs due to the presence of magnetic island. This will lead to temperature flattening in the magnetic island region ~\cite{Fitzpatrick_pop95,Fitzpatrick_pop17,Wangw_pst18}, further affecting the profile of the NTV.
The temperature profile is obtained through the heat diffusion equation~\cite{White_pop15,Teng_NF18},
\begin{center}
\begin{equation}
\hspace{60pt} \frac{3}{2}\frac{\partial(NT)}{\partial t}=\nabla \cdot (\chi_{\mathrm{eff}}\nabla(NT))
\label{eq:T_diff}
\end{equation}
\end{center}
where the initial temperature profile $T_0(r)$ corresponds to the pressure described in Sec.~\ref{sec:non-uni_eq}, and the density is uniform as before. The boundary conditions used here are $T(r=a,t)=T_0(r=a)$, and $T^\prime(r=0,t)=0$. The form of the effective thermal conductivity $\chi_{\mathrm{eff}}$ when including magnetic islands is as follows~\cite{Fitzpatrick_pop95},
\begin{equation}
 \hspace{80pt}\chi_{\mathrm{eff}}(r)= \chi_\parallel g_a g(r)+ \chi_\perp
 \label{eq:chi_eff}
\end{equation}
where $\chi_\bot$ ($\chi_\parallel$) denotes the perpendicular (parallel) thermal conductivity. The Gaussian profile $g(r)= \exp[-(r-r_s)^2/(2g_w^2)]$, the amplitude $g_a=\left(W/4\right)^4 \left(ns/R_0 r\right)_{r_s}^2$ and the standard deviation $g_w$ characterizing the width of the Gaussian profile is proportional to $W$. Here $(\chi_\parallel/\chi_\bot)^{1/4} = 200$ is selected with $\chi_\bot = 1\rm{m^2/s}$ based on experimental experience~\cite{Howell_pop22}. Fig.~\ref{fig:T_flat}(a) and (b) show the typical radial profile of $\chi_{\rm{eff}}$ at the initial moment and the temperature profiles at different time slices with $\beta = 2\%$, respectively. The temperature near the $q=2$ resonant surface shows a significant flattening effect and the temperature gradient is significantly reduced.

\begin{figure}[htbp]
\centering
  \subfigure{\includegraphics[width=0.9\textwidth,angle=0]{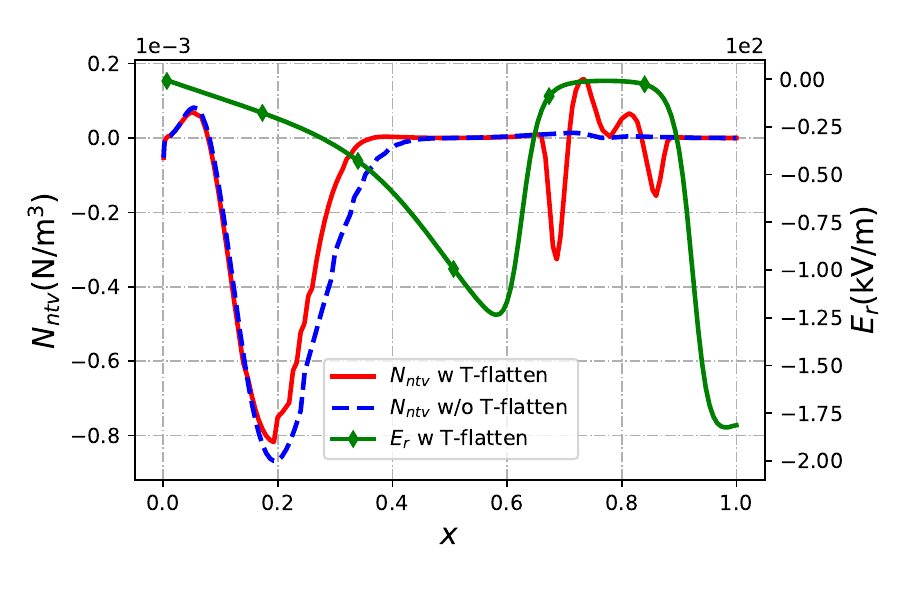}}
  \put(-210,220){(a)}

  \subfigure{\includegraphics[width=0.9\textwidth,angle=0]{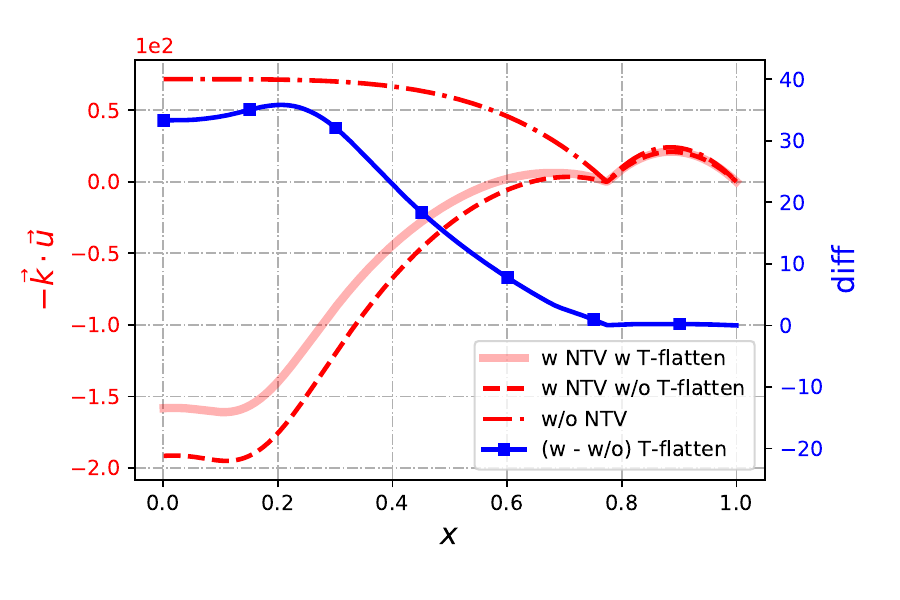}}
  \put(-100,225){(b)}

  \caption{Radial profiles of $N_{\rm ntv}$ (with flattening: red solid; without flattening: blue dashed) and $E_r$ (with flattening: green solid with diamond markers) in the steady state at $\beta = 2\%$ (a); Radial profiles of $-\mathbf{k}\cdot\mathbf{u}$ with NTV and flattening effect (red solid), with NTV but without flattening effect (red dashed), without NTV (red dash-dotted), and the difference between the first two (blue solid with square markers) (b). Here $S=3\times10^5$, $Pr_m=40$, $\Omega_0=2\times 10^2 \mathrm{rad/s}$, $W_C/a=0.292$.}
  \label{fig:locked_1_dpdr_T_flat}
\end{figure}

Now we investigate the model incorporating the temperature evolution equation Eq.~(\ref{eq:T_diff}), with a focus on the impact of the temperature flattening effect. Within the magnetic island region, the NTV amplitude is more pronounced compared with the case without the temperature flattening effect, exhibiting a characteristic of fluctuating distribution. This behavior is associated with the $b_{nm}$ structure (Fig.~\ref{fig:psi_bnm_Tntv_ie}(b)), the misalignment of the peak positions of ion and electron NTV in the radial direction, as well as the radial electric field $E_r$. Local maxima of NTV exist approximately at the position where the $E_r$ vanishes~\cite{Cole_PRL11,Cole_pop11,Satake_PPCF11,Satake_NF13,Honda_NF15}. The effect of temperature flattening on $-\mathbf{k}\cdot\mathbf{u}$ is reflected in the blue line in Fig.~\ref{fig:locked_1_dpdr_T_flat}(b):  the $q=2$ surface remains unchanged, whereas the core flow exhibits the most significant variation. Similar to the previous discussion, the temperature flattening effect changes the local distribution of NTV, which directly affects the $\Delta\Omega_\phi$ profile and indirectly affects the $\Delta\Omega_\theta$ profile. However, the finally superposed $-\mathbf{k}\cdot\mathbf{u}_s$ shows almost no change, indicating that the locked state is not altered.

\begin{figure}[htbp]
\centering
  \subfigure{\includegraphics[width=0.9\textwidth,angle=0]{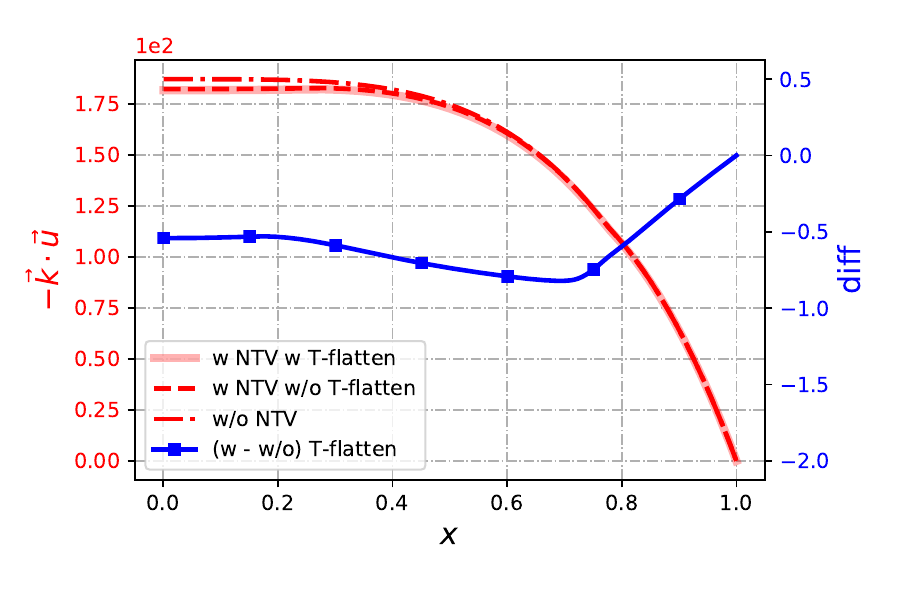}}

  \caption{ Radial profiles of $-\mathbf{k}\cdot\mathbf{u}$ with NTV and flattening effect (red solid), with NTV but without flattening effect (red dashed), without NTV (red dash-dotted), and the difference between the first two (blue solid with square markers). Here  $S=2.44\times10^3$, $Pr_m=1$, $\Omega_0=2\times 10^2\mathrm{rad/s}$, $W_C/a=0.146$.}
  \label{fig:unlocked_1_dpdr_T_flat}
\end{figure}

Fig.~\ref{fig:unlocked_1_dpdr_T_flat} presents the results of the unlocked case with $\beta = 2\%$. It is found that the introduction of the temperature flattening effect does not change the original mode state, i.e., the unlocked mode state. Whether the temperature flattening effect reduces (Fig.~\ref{fig:unlocked_1_dpdr_T_flat}) or elevates (Fig.~\ref{fig:locked_1_dpdr_T_flat}) the global parallel flow depends on the characteristics of the NTV profiles, and there is no straightforward linear correlation between them.

\section{Summary and discussion}
\label{sec:sum}
In summary, the plasma flow evolution in a cylindrical tokamak responding to single-helicity RMP considering the NTV effect has been predicted and evaluated using a single-fluid theoretical model. This model takes into account the viscous torque, the resonant EM torque induced by the response of plasma to RMP, and the NTV torque. Representative cases of both ``locked'' and ``unlocked'' states of the parallel plasma flow, along with the steady states of the nonlinear response, have been considered and reported, with separate discussions conducted on the two scenarios: uniform pressure and non-uniform pressure. Under non-uniform pressure condition, we have also solved the model that incorporates the temperature equation, and discussed the temperature flattening effect in the magnetic island region as well as its influence on the NTV torque and plasma rotation.
Calculations show that the introduction of the NTV effect does not alter the original locked or unlocked state. Under the parameters adopted in this paper, NTV effect barely affects the evolution of magnetic island width for the uniform pressure case. However, it reduces the $\Omega_\phi$ and $\Omega_\theta$ rotation in the core region, as well as the parallel flow $-\mathbf{k}\cdot\mathbf{u}$. For the non-uniform pressure case, the global magnitude of $N_{\rm{ntv}}$ increases as $\beta$ increases; this trend slightly slows down the parallel flow $\omega_s$ at the resonant surface and reduces the phase difference $\varphi$. However, the evolution of the magnetic island width $W$ is almost unaffected by $\beta$, which ultimately decreases the magnitude of $N_{\rm{mag}}$. Nevertheless, these two driving terms collectively maintain the original locked mode state. As $\beta$ increases, the parallel flow $-\mathbf{k}\cdot\mathbf{u}$ in the core region may decrease or even increase in the opposite direction. In addition, the temperature flattening effect does not alter the original locked or unlocked mode state.

The neoclassical poloidal viscosity (NPV) damps the poloidal flow in tokamaks and exerts a pronounced influence on plasma poloidal rotation~\cite{Gianakon_pop96,Gianakon_pop02,Taguchi_pop13,Jepson_pop21,Gustafson_pop24}. Therefore, the NPV effect should also be taken into account in subsequent studies. Another promising development direction is to couple the NTV effect to more sophisticated and realistic nonlinear MHD models, such as NIMROD~\cite{Sovinec_JCP04}. Beyond the resistive single-fluid model, two-fluid and kinetic effects, as well as 2D and 3D neoclassical effects (including ion polarization drift and turbulent transport effects) in presence of finite island width, are all essential to account for additional modifications to the Rutherford equation for island growth and saturation, such as those from the neoclassical tearing mode, in the finite-$\beta$ regime of tokamak plasma response to RMPs.

\ack
The fruitful discussions and valuable comments with participants at the 16th Asia Pacific Physics Conference are appreciated. Fangyuan Ma thanks Drs. Wenlong Huang, Haolong Li and Xingting Yan for helpful discussions. This work is supported by the National MCF Energy R\&D Program of China under Grant No.~2019YFE03050004. The computing work in this paper is supported by the Public Service Platform of High Performance Computing by Network and Computing Center of HUST.

\newpage
\bibliographystyle{iopart-num-nourl}
\bibliography{myref}

\newpage

\clearpage

\clearpage

\clearpage

\clearpage

\clearpage

\clearpage

\clearpage

\end{document}